\documentclass{article}

\usepackage[a4paper, total={8in, 10in}]{geometry}
\usepackage{graphicx}%
\usepackage{multirow}%
\usepackage{amsmath,amssymb,amsfonts}%
\usepackage{amsthm}%
\usepackage{mathrsfs}%
\usepackage[title]{appendix}%
\usepackage{xcolor}%
\usepackage{textcomp}%
\usepackage{manyfoot}%
\usepackage{booktabs}%
\usepackage{algorithm}%
\usepackage{algorithmicx}%
\usepackage{algpseudocode}%
\usepackage{listings}%
\usepackage{subcaption}%
\usepackage{mathtools}%
\usepackage[T1]{fontenc}
\usepackage[utf8]{inputenc}
\usepackage{authblk}
\usepackage{url}
\usepackage{hyperref}
\usepackage{tabularx}
\begin{document}
	
	\title{A Scalable Cloud-Orchestrated and Service-Oriented Multi-Domain QKD Network with PQC Integration}
	
	\author[1,2]{Konstantinos Krilakis}
	
	\author[2]{Antonia Tsili}
	
	\author[1,2]{Aikaterini Mandilara}
	
	\author[1,2]{Dimitris Syvridis}
	
	\affil[1]{Department of Informatics and Telecommunications, National and Kapodistrian University of Athens}
	
	\affil[2]{Eulambia Advanced Techonlogies Ltd.}
	
	\date{}

	\maketitle
	\abstract{Quantum key distribution (QKD) offers unconditional security but existing QKD networks remain difficult to scale across heterogeneous infrastructures and administrative domains due to vendor-specific interfaces, trusted-node constraints, and limited interoperability. This work presents a flexible multi-domain and multi-site quantum-secure network architecture integrating vendor-agnostic QKD, SDN orchestration, and cloud-managed trust services. Communication is based on Zero Trust Network Access protocols featuring multi-level authentication mechanisms building upon post-quantum cryptography (PQC) signature and key encapsulation algorithms. The system is deployed on a real-world testbed with domains incorporating QKD nodes from 3 vendors, as well as domains without QKD infrastructure elements. Experimental results show that PQC and SDN overhead remain relatively low even on constrained devices, with the main bottleneck being QKD key retrieval and vendor-specific key streaming limitations. The proposed framework extends quantum-safe key transport beyond native QKD boundaries while preserving flexibility, interoperability, and compatibility with existing infrastructures.}
	
	\section{Introduction}
	Research in applied cybersecurity is essentially converging towards two emerging frontiers~\cite{chamola_information_2021} that offer the properties required to safeguard the global network against the dangers of quantum computer cyber attacks; quantum key distribution (QKD) and post-quantum cryptography (PQC). Each approach entails distinct trade-offs. QKD requires specialized and often costly hardware, but can offer information-theoretic security (ITS)~\cite{RevModPhys_92_025002} under well-defined assumptions. PQC can be easily integrated into existing infrastructures, but misses an equilibrium that balances proven security guarantees and performance efficiency.
	
	Quantum Key distribution is a method for distributing cryptographic keys with theoretically unconditional security. It operates at the physical layer of the network and the basic scheme includes two terminals (Alice and Bob) connected with an ideally dark fibre of limited length. Clients on either Alice or Bob end ultimately receive symmetric cryptography keys through authorised and secure access to the corresponding QKD terminal. This pairwise scheme substantially restricts application range and scalability, especially under the constraint of the devices' increased financial cost. Under certain assumptions and configurations, symmetric keys can be shared between clients on the outer nodes of pairs that are connected through chain-like QKD links. The links are formed using trusted nodes (TNs)~\cite{salvail_security_2010}, and extend the reach of key agreement beyond the pair's distance limitations. Nevertheless, installation of multiple QKD nodes increases the cost even more, and the symmetric keys can only be shared between clients attached to nodes that are parts of this sequential chain at a certain rate. Optical switches could overcome the issue of strict sequential interactions~\cite{yang_all_2021}, as they offer flexible path reconfiguration, but they introduce additional insertion loss at every switching stage, reducing the secret key rate and maximum achievable distance. Furthermore, optical switches only provide passive routing and do not regenerate or store keys, meaning losses and errors propagate across the entire path. In contrast, trusted nodes terminate the quantum link at each hop, generate fresh keys, and forward them using one-time-pad encryption, enabling longer distances and more stable performance. As a result, trusted-node architectures remain more practical for large-scale QKD deployments.
	
	Several works over the past years have focused on overcoming QKD limitations ranging from theoretical key management and new hybrid protocols that improve the trust over relay nodes~\cite{pietri_quantum_2025} to practical QKD network (QKDN) extensions~\cite{chen_implementation_2021, horoschenkoff_demoquandt_2025, chen_implementation_2025}. Dervisevic et al.~\cite{dervisevic_quantum_2025} conducted a survey providing a comprehensive system-level overview of key management in QKD networks, focusing on architectures based on trusted relay nodes. The authors highlight the differences between QKD and classical key management, emphasizing the need to decouple key generation and consumption due to the limited rate and point-to-point nature of QKD links. The survey identifies open challenges such as scalability, synchronization, and multi-vendor interoperability. Works inspired by the decoupling of key generation and consumption propose the integration of key pools, for example Wang et al.~\cite{wang_quantum_2021} introduce an algorithm based on balancing key resources and routing hops by managing resource consumption and QKD key accumulation in pools, in the context of a multi-domain network. Zhu et al.~\cite{zhu_qkd_2024} discuss QKD key provisioning to partially-QKD-deployed optical networks using multi-level key pool slicing. The goal is to efficiently accumulate and distribute key material to E2E applications either they have access to key streams of QKD nodes or not, a useful and detailed key management approach which requires extensive configurations, thus a laborious transition of existing QKDNs, which still need to undergo significant scaling~\cite{horoschenkoff_demoquandt_2025} challenges. Another set of works have concentrated on the fact that authentication in QKD interactions is not standardised, attempting to efficiently and securely inject PQC methods to this aspect. Kozlovi\v{c}s et al.~\cite{meng_quantum_2023} propose two protocols supporting the key agreement between two clients that obtain QKD-produced keys, while upgrading TLS 1.3 with PQC algorithms. The idea is that the clients receive parts of the keys from different pathways, further securing key distribution, with servers acting as the interface of the QKD terminals. Atutxa et al.~\cite{atutxa_authentication_2025} employ PQC digital signatures and study their effect on Quantum Bit Error Rate (QBER), which measures the ratio of incorrect bits to total bits received, acting as an indicator of eavesdropping or channel noise, while Tsili et al~\cite{tsili2025} create an automated PQC PKI framework that can be deployed on QKDNs. 
	Finally, an applied research direction of interest introduces real QKDNs deployments with PQC integration consisting multiple nodes and TNs~\cite{martin_madqci_2024, mendez_sdn-based_2024, mendez_switching_2026}. Such solutions succeed in demonstrating real-world application and viable adaptations with tangible results -- which is the aim behind the current work. Recent efforts have also explored federation mechanisms for extending QKD infrastructures beyond isolated metropolitan deployments. In particular, Barral et al.~\cite{barral_interconnecting_2026} proposed an ETSI-compliant hybrid PQC/QKD federation architecture for interconnecting geographically separated QKD ``islands'' through distributed key management overlays and trusted relay federation. Their approach combines QKD-generated secrets with PQC mechanisms to enable secure inter-domain key transport across heterogeneous infrastructures while maintaining interoperability with ETSI GS QKD standards. Similar to MadQCI, the work emphasizes multi-vendor interoperability, distributed orchestration, and practical deployment considerations in carrier-grade environments. However, the federation model remains primarily centered around KMS-level relay coordination and ETSI transport overlays. A detailed breakdown of the corresponding specifications can be found in the Supplementary Information. 
	Here, we present a multi-site enterprise-level network infrastructure and dedicated cloud management designed to enable quantum-secure communication among all network entities (NEs). Contrary to Mendez et al.~\cite{mendez_sdn-based_2024}, where PQC is used in some segments of a QKDN for emulating long-distance QKD connections, post-quantum methods are employed to securely conduct the provision of secret key material that could be provided via QKD or other means, and other cryptographic services, over classical channels. As a result, the need for extensive QKDN installations is avoided, while exploiting existing QKD pairs and essentially extending quantum-safe key relay in a way compatible to legacy networks. The architecture is organized into separate, autonomous domains with a central supervisor which provides the capability of the system is the delivery of cryptographic keys generated via QKD to NEs across domains, as well as the creation of secure tunnel links for encrypted data exchange. More specifically, we distinguish the following properties and novelties:
	
	\begin{figure}[tb!]
		\centering
		\includegraphics[width=.7\textwidth]{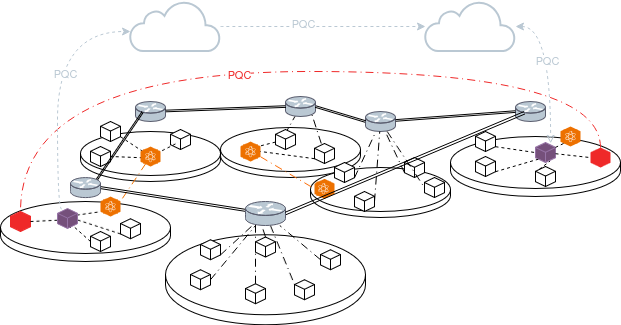}
		\caption{\textbf{High-level schematic illustration of the proposed network model.} Each disk represents a domain; The white cubes represent client NEs; the \textcolor{purple!60!blue}{purple} cubes represent Domain Secure Services Providers (DSSPs); the client application communication with PQC is shown in \textcolor{red}{red}; the secure services PQC connections are shown in \textcolor{gray}{gray}.}
		\label{fig:overview}
	\end{figure}

	\begin{itemize}
		\item A native, cross-platform, multi-domain network architecture with support for heterogeneous topologies.
		\item An agnostic design supported by a zero-touch provisioning (ZTP) driver framework.
		\item Integration of multi-vendor QKD systems within some domains, combined with PQC KEMs to enable unlimited distance quantum-secure key relay.
		\item A unified security layer incorporating multi-factor authentication and cloud-based key management for trust, identity, and interoperability.
		\item SDN-driven automation with detailed telemetry for monitoring and reporting.
	\end{itemize}
	
	The above network was experimentally evaluated on a 12-domain testbed with heterogeneous hardware, quantifying end-to-end setup latency and	packet processing overhead. A high-level illustration of the infrastructure is depicted in Figure~\ref{fig:overview}, where each disk represents a domain comprising NEs, a gateway router, a Domain Secure Services Provider (DSSP) and some domains integrate a QKD node as the key generator. The figure shows two kinds of PQC-protected interations; the direct communication of two NEs located in different domains (red), established through the service provision of the DSSP of each domain, and the management of the DSSPs from the cloud (gray), as cloud-based routers facilitate inter-domain connectivity and aggregation of the domains orchestrated by the visualised-SDN data plane components.
	
	The remainder of this work is organized as follows. In Section~\ref{sec:overview}, we present the resulting multi-domain quantum-secure network architecture and describe its underlying classical and quantum infrastructure. We detail the system operation and the workflows enabling inter-domain key relay across hybrid quantum-classical environments, but also list experimental results derived from our heterogeneous 12-domain deployment. In Section~\ref{sec:components}, we describe the logical organisation of our system and the main components developed for the construction and seamless operation and orchestration of the offered cybersecurity services. Finally, we analyse the practical implications, limitations, and scalability challenges of the proposed approach, concluding with directions for future work towards large-scale quantum-secure deployments.
	
	\section{Network Topology}\label{sec:overview}
	
\begin{figure}
    	\hspace{1cm}
    	\begin{subfigure}[t]{0.25\textwidth}
    		\centering
    		\includegraphics[width=\textwidth, trim={0.5cm 1cm 1cm 1cm}, clip]{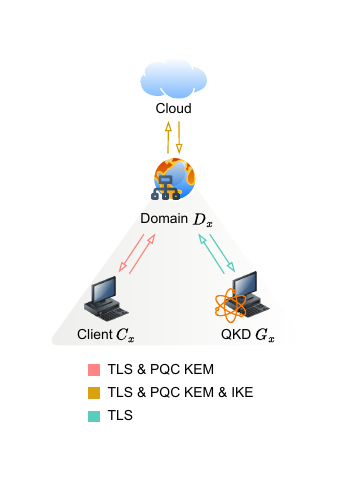}	
    		\caption{A domain with a QKD terminal as key generator and the authentication mechanisms per connection type according to the participating peers.}
    		\label{fig:domain-qkd}
    	\end{subfigure}
		~
		\begin{subfigure}[t]{0.6\textwidth}
			\centering
			\includegraphics[width=\textwidth, trim={0.5cm 0.5cm 1cm 0cm}, clip]{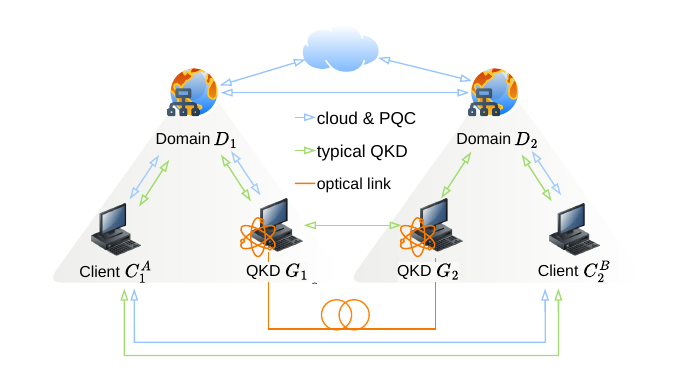}
			\caption{A scenario wherein the users belong to different quantum domains and the QKD nodes form an Alice-Bob pair. The \textcolor{green!20!gray}{green} arrows represent conventional QKD usage. The \textcolor{blue!50!white}{blue} arrows represent QKD-derived key sharing through forwarding PQC encapsulated messages.}
			\label{fig:interdomain}
		\end{subfigure} 
		\caption{\textbf{Intradomain operation and a basic interdomain communication scenario.} The figures illustrate the architecture and message exchanges within a single domain, together with its connectivity to a neighbouring domain through both classical and quantum channels. The intradomain workflow presents the service transactions and corresponding authentication mechanisms in a simplified quantum domain comprising a client, a Domain Secure Services Provider (DSSP), and a QKD node acting as the key generator. The interdomain scenario depicts the conventional Alice–Bob QKD configuration, where the two QKD nodes form a direct QKD pair.}
	\end{figure}

	The proposed work introduces a multi-site enterprise-grade network infrastructure that allows \textit{quantum secure} communication between any network entities (NEs) involved. Our deployment consists of twelve (12) independent network domains, designed after real-world network equivalents, to which NEs and key generators are attached, see for example Figure~\ref{fig:domain-qkd}.
	Formally, an infrastructure domain is defined as a physical or virtual network segment --implemented through hardware NAT boundaries, VLAN isolation, or equivalent mechanisms-- that exposes a functional grouping of network resources governed by a specific set of security policies, authentication protocols, and distributed administration rules. The domains' inner topology is abstracted to a version that can be matched to a wide range of fundamental network units with little to no requirements. Each domain contains a router that functions as a gateway, a low-power device functioning as the Domain Secure Services Provider (DSSP), and at least one key generator and one client. Clients are organised in client pools, where a client pool is a virtualized SDN network pool that combines flow rules and port policies policies to create a programmable network under each hardware domain. We focus on the application wherein the key generators are QKD terminals, thus we distinguish two kinds, the classical or \textit{non-quantum}, and the QKD or \textit{quantum} domains. Classical domains adhere to classical channel communication principles. Quantum domains contain a QKD node each and are vendor-agnostic. 
	
	\begin{figure}
		\centering
		\includegraphics[width=\textwidth]{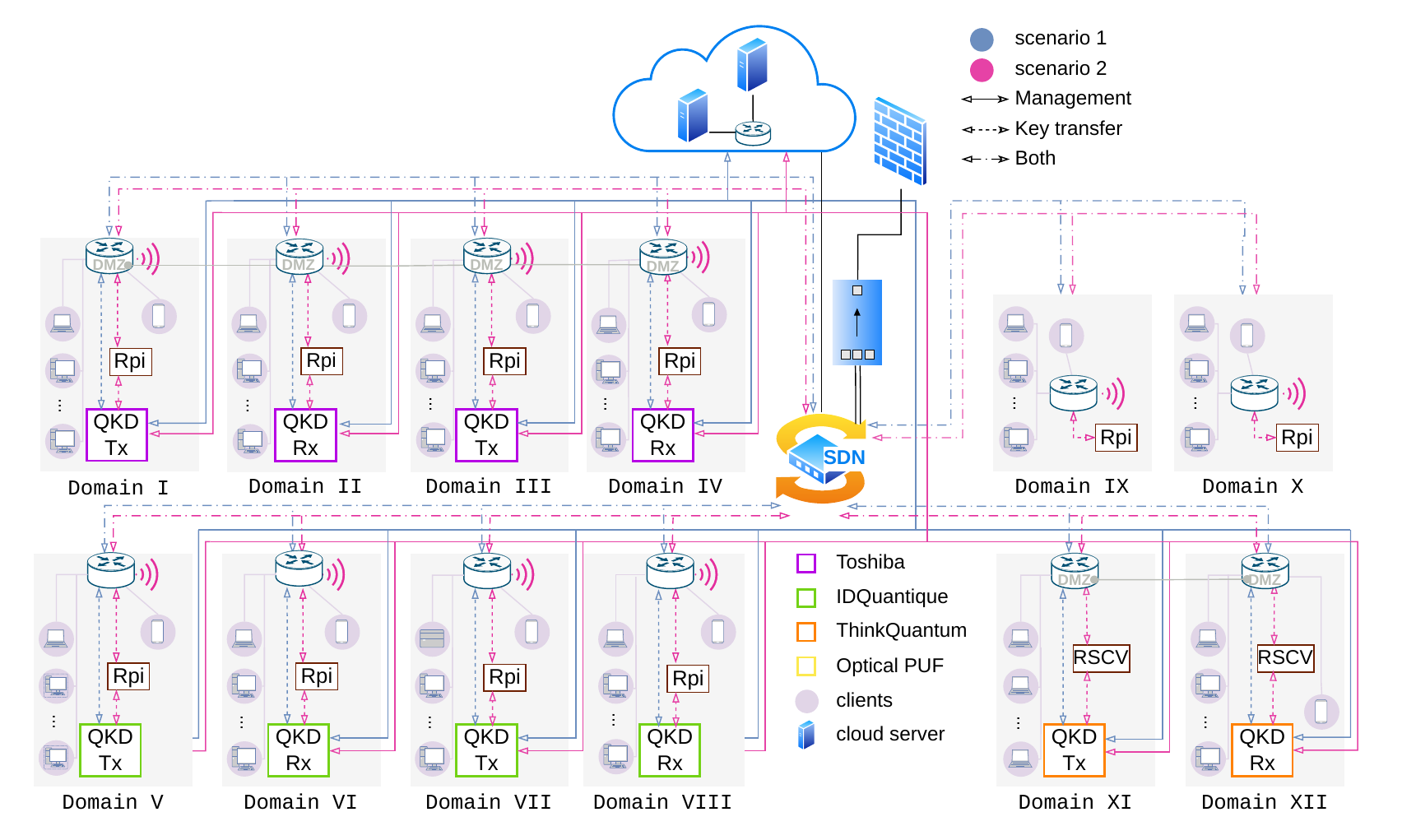}
		\caption{\textbf{Detailed overview of the testbed's architecture.} This figure shows explicit connections within the domains and the cloud. Dotted arrow links show key transfer within each domain, solid arrows show management control originating from the cloud, and the remaining interdomain arrows are governed by both connection types. There are three (3) possible scenarios concerning intradomain key transfer, depending on transaction type: the \textcolor{blue!40!gray}{blue} scenario (1) describes usual QKD function; in the \textcolor{magenta}{magenta} scenario (2), QKD keys are processed from the gateway before reaching the client; the third scenario uses a wifi link between the gateway and the client. }
		\label{fig:topology}
	\end{figure}
	
	\subsection{Classical Network}
	
	Each administrative domain is defined by minimal design assumptions enabling a technology-agnostic, service-oriented deployment that simplifies management and enhances monitoring. Domains are delimited by hardware gateway routers providing local connectivity, including wireless access where required, while domain environments are realized using heterogeneous computing platforms ranging from constrained Raspberry Pi and Android-powered devices to x64 and RISC-V-based systems. These devices support gateway operations and services deployed within demilitarized zones (DMZs) in conjunction with the router. Domain components interconnect through a combination of VLAN-based traffic isolation, proprietary VPN tunnelling, and identity-based segmentation orchestrated by the cloud control plane, whereas clients connect to the domain router either directly through Wi-Fi links or virtually through software-defined networking (SDN). The SDN infrastructure employs OpenFlow 1.3~\cite{McKeown2008OpenFlowEI} through an ONOS controller~\cite{berde2014onos} hosted within the cloud environment and coordinated with a dual-mode managed switch running in virtualized SDN mode and supporting both aggregated (non-SDN) and SDN-based architectures on the same device. Network Address Translation (NAT) is used to obscure internal topology and retain traffic processing within the SDN data plane, while the SDN switch dynamically redirects traffic according to controller-defined flow rules and port policies, effectively placing clients into different logical positions across segmented network domains. Consequently, each NE can be programmatically assigned to dedicated configuration and operational positions, enabling dynamic establishment and orchestration of domain layouts, as depicted in Figure~\ref{fig:topology}. For simplicity, the topology assumes one Network Interface Card (NIC) per device, although multiple devices support concurrent multi-NIC parallel connections. The network is protected by using a open source Network Access Control and Identity Manager, behind a 1G hardware firewall. This consolidates control and monitoring functions while enforcing security boundaries over different external testbeds using Media Access Control Security (MACsec) and Internet Protocol Security (IPsec) connections, creating a federated enterprise network. Our implementation remains fully independent of the OQS project and relies exclusively on portable standardized algorithm releases, ensuring portability and avoiding tight coupling with specific hardware platforms while providing reproducible measurements. Working in parallel with existing directory-service models, the system is deployed as an additional non-discoverable hidden directory-service layer coordinated through compatible upper management levels. Furthermore, cryptographic functionality is implemented through a lightweight custom application-layer library with minimal kernel dependencies (e.g., KTLS), rendering the software transferable, transparent, and readily integrable with existing applications.
	
	\subsection{QKD Network}
	At the lowest level and closest to the physical characteristics of the system lies the quantum layer, which is largely driven by the configuration of the QKD nodes. The quantum layer comprises pairs of QKD nodes running a vendor-specific version of the BB84 protocol~\cite{bennett1984quantum}, among which key agreement takes place in the quantum channel, and pairs that form TNs. TNs operate as repeaters in QKD networks, extending the physical reach of the key agreement process and creating a single end-to-end logical chained link between QKD nodes. This functionality requires that the QKD devices forming the TN are purposely configured for this reason, while remaining in a protected environment. From an operational perspective, each QKD node is associated with a specific administrative domain managed by a DSSP, which acts as a trusted proxy for inter-domain key relaying. When forwarding a key $K$ to a neighbouring domain, the DSSP retrieves a locally generated QKD key $K'$ from the quantum layer and computes $K \oplus K'$. The resulting value is then transmitted to the next domain, where the corresponding DSSP can recover $K$ using the shared key material. In this way, secure end-to-end key transport across multiple administrative domains is achieved without directly exposing the original key during transit. Here, we assume the existence of inter-domain links that lack the interoperability and protection mechanisms required for seamless trusted-node operation, particularly in multi-vendor deployments. We refer to such incomplete relay links as dangling chain edges (DCEs). The chain segments between DCEs are reinforced with PQC KEMs, thus filling the chain's non quantum-resistant gaps and yielding a chained link of arbitrary length. We exhibit this paradigm on a testbed of five (5) QKD pairs, two (2) Toshiba MU~\cite{Toshiba_MU_QKD} pairs, two (2) ID Quantique Clavis XG~\cite{IDQ_ClavisXG} pairs and a ThinkQuantum QUKY~\cite{ThinkQuantum_QUKY} pair. Photographs of the physical devices can be found in the Supplementary Information.
	
	\subsection{TN emulation}
	We implement an emulated equivalent of a conventional TNs via a point to point topology, which can further safeguard the communication of DCEs, provided that a MACsec tunnel has been previously setup. The tunnel can be deployed either as cloud-hosted Secure Access Service Edge (SASE) endpoints or as static on-premises endpoints. In both cases, the endpoint service embeds Identity and Access Management (IAM) and Network Access Control (NAC) functionalities through a RADIUS-based authentication and authorization interface, allowing centralized enforcement of access-control policies independently of the underlying network infrastructure. The endpoint service operates a PQC-KEM-secured post-quantum mutual TLS streaming encryptor/decryptor pair that accepts only authorized DSSPs as peers. Each pair is instantiated with dynamically defined protocol parameters, dedicated PQC KEM root keys, and post-quantum mTLS certificates, making every connection uniquely bound and non-interchangeable. A schematic of the proposed setup can be found in the Supplementary Information.
	
	\subsection{Encryptors}
	The DSSP of domains 1, 2, 3, 4, 9, and 10 have integrated software encryptors that are cross-platform, crypto-agile packet encryptors designed to deliver high-assurance tunneling services between network domains. These software services support multi-domain interoperability and are engineered to operate within heterogeneous cryptographic environments, providing multi-layer encryption, identification through PQC key pairs and multi-vendor QKD support. Configuration is performed through a PQC-KEM-secured network management interface, which communicates directly with the cloud key management system (KMS). This mechanism ensures post-quantum-resilient provisioning of configuration parameters and cryptographic material.
	The encryptors support different protocols for key-generating sources, including:
	\begin{itemize}
		\item QKD ETSI 014~\cite{ETSI014}
		\item SKIP
		\item Direct integration with the cloud KMS as a key-provisioning entity
	\end{itemize}
	This flexibility enables seamless operation in hybrid classical-quantum key distribution environments.
	
	\section{System Operation}\label{sec:system} 
	
\begin{figure}
		\centering
		\includegraphics[width=.5\textwidth, trim={0.7cm 1cm 0cm 0.5cm}, clip]{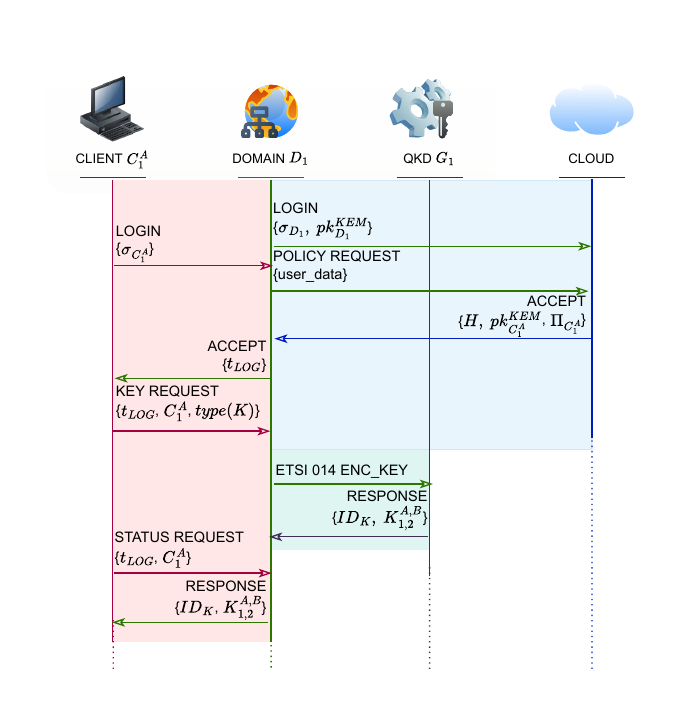}
		\caption{\textbf{The intradomain transactions leading to key generation.} The client logs onto the domain services, requesting a key from the QKD terminal. Coloured areas denote the authentication protocols: \textcolor{red!35!white}{pink} for TLS and KEM public key identification; \textcolor{blue!35!white}{blue} for TLS, KEM public key identification and Internet Key Exchange (IKE); \textcolor{green!20!gray}{green} for single-direction or mutual TLS. }
		\label{fig:keygen-trans}
	\end{figure}

\begin{figure}
		\centering
		\includegraphics[width=.7\textwidth, trim={0.7cm 1cm 0.8cm 0.5cm}, clip]{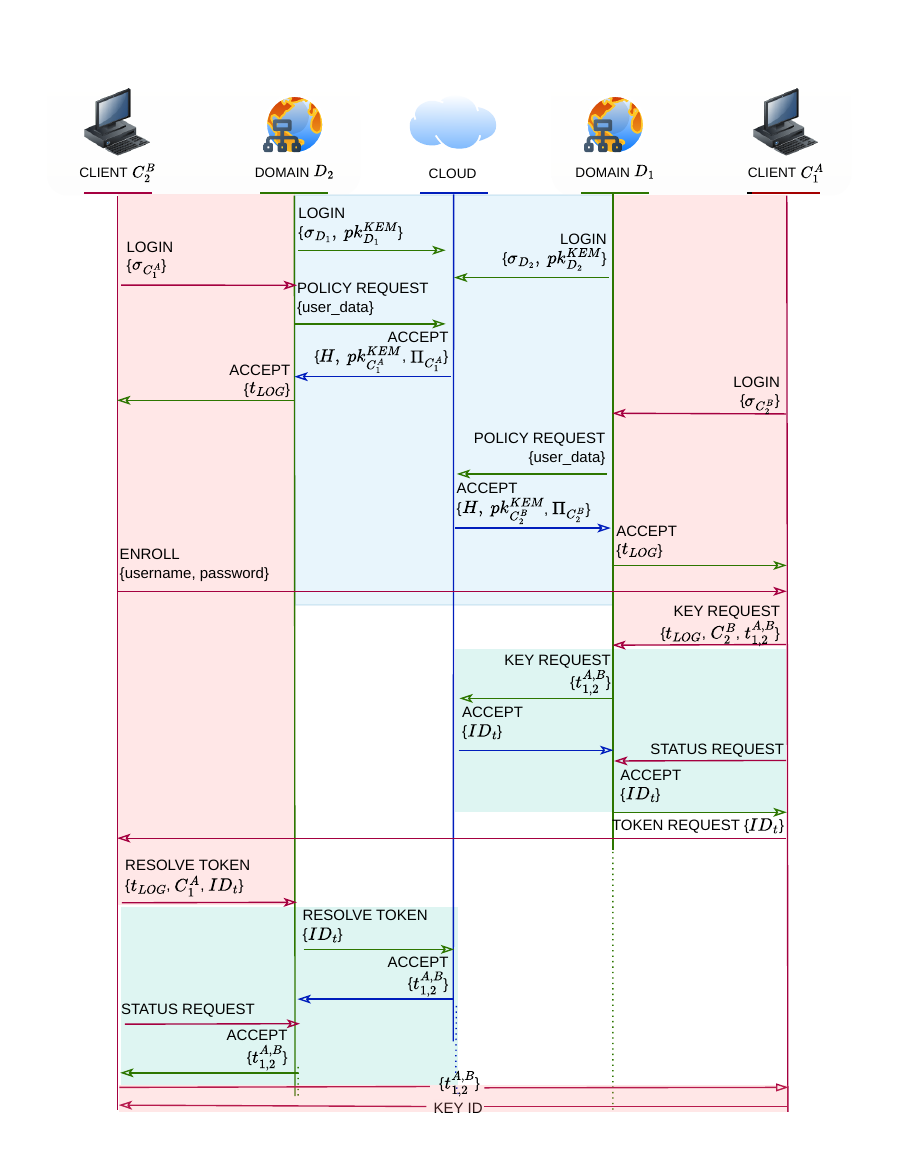}
		\caption{\textbf{The interdomain transactions for key agreement between clients $C^A_1$ and $C^B_1$.} The first login steps are the same as in Figure~\ref{fig:keygen-trans}. Then, $C^B_2$ enrolls to the services exposed by $C^A_1$ and $C^A_1$ stores a token on the cloud. The peer receives and resolves the token, which is used to obtain the ID of the symmetric key to be used for encryption. Coloured areas denote the authentication protocols (see Figure~\ref{fig:keygen-trans}).
		}
		\label{fig:relay-trans}
	\end{figure}

	Consider the setting where a set of $m$ NEs designated as clients $C^1, C^2, ..., C^m$ is distributed among domains $D_1, D_2, ..., D_n$. Domains $D_x\, x\in [n]$ integrating a QKD node $G_x$ are hereafter called quantum domains, and are denoted $D^*_x$ if they fall into the DCE category (see Section~\ref{sec:overview}). 
	The assignment yields 
	\begin{gather*}
		C^{y_1}_1, C^{y_2}_1, ..., C^{y_{p_1}}_1 \text{ of } D_1, \\
		C^{y_{p_1+1}}_2, C^{y_{p_1+2}}_2, ..., C^{y_{p_1+p_2}}_2 \text{ of } D_2 \\
		\vdots
	\end{gather*}
	where $y_i \in [m]$, $i \in \{1,2,...,p_1, p_1+1, ..., p_1+p_2, ... \}$. Let us denote $C^A_1 \coloneqq C^{y_1}_1$ after `Alice', and $C^B_2 \coloneqq C^{y_{p_1+1}}_2$ after `Bob'. We assume that client device $C^A_1$ of domain $D_1$, which comprises key generator $G_1$, requests cryptographic material that will allow the encrypted communication with client $C^B_2$ of domain $D_2$, as in Figures~\ref{fig:interdomain}-\ref{fig:interdomain-nonqkd}. It follows that, the ultimate goal for $C^A_1$ and $C^B_2$ is to share the symmetric key $K^{A,B}_{1,2}$ for encryption and decryption. During the domain and client pool creation, each NE is provided with digital certificates and an initial key pair $(pk^{KEM}, sk^{KEM})$ produced with a PQC KEM, bound to its identity (see Section~\ref{sec:key-management}). The public key $pk^{KEM}$ is stored on the cloud for peer-to-peer identification and trust establishment. 
	
	Intradomain modules and services are consistently accessible within each domain, maintaining their integrity while forming functionally complete units that can operate independently and collaborate seamlessly. Inside the domain $D_1$, a client $C^A_1$ is authenticated through multi-factor means. Namely, successful authentication is dependent on TLS and an additional security layer based on the ability of $C^A_1$ to decrypt with their secret key $k^{KEM}_{s}$, confirming their identity. Once the client device $C^A_1$ has been authenticated, an encapsulated key request is sent to the domain gateway. The domain services of $D_1$ will attempt to gain authenticated access to the cloud services, and retrieve the updated policy restrictions that apply to client $C^A_1$. Depending on the nature of the request, the client will proceed to perform a key agreement with client $C^B_i, \ i\in [n]$, or the domain services will prompt the local key generator $G_1$ to provide key $K^{A,B}_{1,2}$. 
	
	In more detail, suppose that client with identity profile $C^A_1$ wants to request a QKD-produced key $K^{A,B}_{1,2}$ from their local QKD terminal. The client logs onto the domain services using their signature $\sigma_{C^A_1}$ and the respective user-driven policy $\Pi_{C^A_1}$ that must be followed is updated. The update is performed by interchange of the domain-facing and cloud services, wherein the client destination is identified by their PQC KEM public key $pk^{KEM}_{C^A_1}$. If authorised, the domain-facing services allow role-based access (RBAC) and produce a token $t_{LOG}$. The client subsequently uses their profile $C^A_1$ and the received token $t_{LOG}$ to request and receive the QKD-produced key $K^{A,B}_{1,2}$, along with its identifier $ID_K$, through the domain services. The process is shown in full in Figure~\ref{fig:keygen-trans}. Next, consider the case wherein $C^B_1$ has initiated key agreement with $C^A_1$; cloud key, domain and identity management will perform multi-layer authentication between domain and cloud, similar to the previous-stage authentication mechanism, and will supply $C^B_1$ with the means for retrieving the correct key from the local key generator, a.k.a the QKD node $G_1$.
	
	\section{Connecting DCEs for Multi-domain Key Distribution}\label{sec:multi-domain}
	
	Our vendor-agnostic approach to key relay is based on cloud-managed post-quantum key encapsulation and authentication, as well as SDN functionality, all conveniently complementing and reinforcing QKD across domains. PQC mechanisms are efficiently exploited, enhancing the resilience and forward secrecy of communications, while cloud management ensures that only legitimate and continuously validated users access the provided security services. Ultimately, one key can be shared between two NEs without distance limitations following a two-phase process described below. The handshake phase ensures transfer only to strictly authorized parties, following the Zero Trust Network Access (ZTNA) premises, while the transfer phase defies vendor-specific integration, e.g. as required by interfaces conforming to ETSI GS QKD 015~\cite{ETSI015}.
	
	\subsection{Handshake Phase}
	The functionality described in Section~\ref{sec:system} is applied consistently across all domains. Initially, domains $D_1$ and $D_2$ log onto the cloud services, updating the policies and permissions applicable to their corresponding clients $C^A_1$ and $C^B_2$, and the clients, in turn, log onto the respective domain services. These login steps of this process are the same as in Figure~\ref{fig:keygen-trans} for all endpoints of the respective domains $D_1$ and $D_2$. All clients expose a service API specifically purposed to serve communication intentions. Hence, once clients $C^A_1$ and $C^B_2$ have received a respectively unique login token $t_{LOG}$ for service access, $C^B_2$ enrolls to the services exposed by $C^A_1$. Client $C^A_1$ will send its profile data, the login token and a token $t^{A,B}_{1,2}$ to the cloud through the domain services.  Client $C^A_1$ stores token $t^{A,B}_{1,2}$ on the cloud, in order to be shared with the communication peer, and expects to receive it from $C^B_2$ as proof that it belongs to a known, authenticated, and accessible domain, in this case $D_2$. Afterwards, $C^A_1$ will communicate the identifier $ID_t$ to $C^B_2$. The latter, once authorized, will resolve the token $t^{A,B}_{1,2}$ with identifier $ID_t$ by accessing the cloud through the domain services, then encapsulate and present $t^{A,B}_{1,2}$ to $C^A_1$, thereby establishing trust. 
	
	\begin{figure}
		\centering
		\includegraphics[width=.7\textwidth,trim={0.5cm 0.45cm 1cm 0cm}, clip]{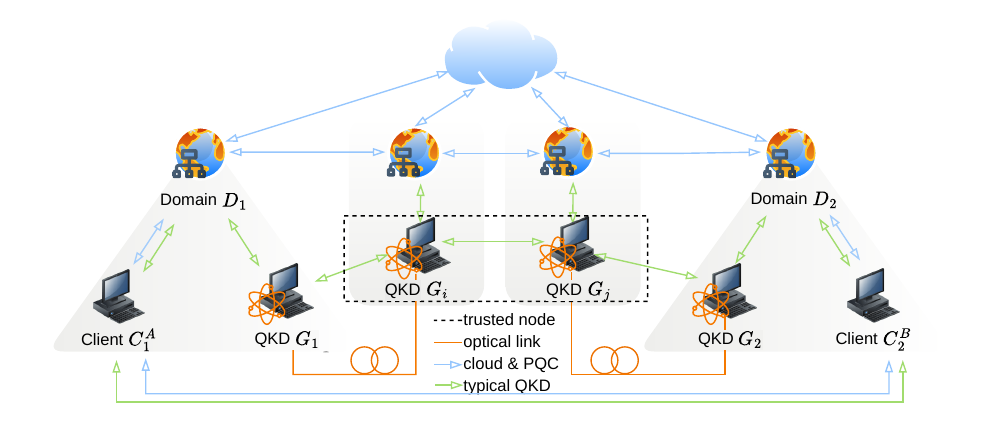}
		\caption{\textbf{The clients are placed in two remote quantum domains with QKD terminals connected through a TN.} The \textcolor{green!20!gray}{green} arrows represent the conventional QKD usage. The \textcolor{blue!35!white}{blue} arrows represent QKD-derived key sharing through forwarding PQC encapsulated messages.}
		\label{fig:interdomain-tn}
	\end{figure}
	
	\begin{figure}
		\centering
		\includegraphics[width=.7\textwidth,trim={0.5cm 0cm 1cm 0cm}, clip]{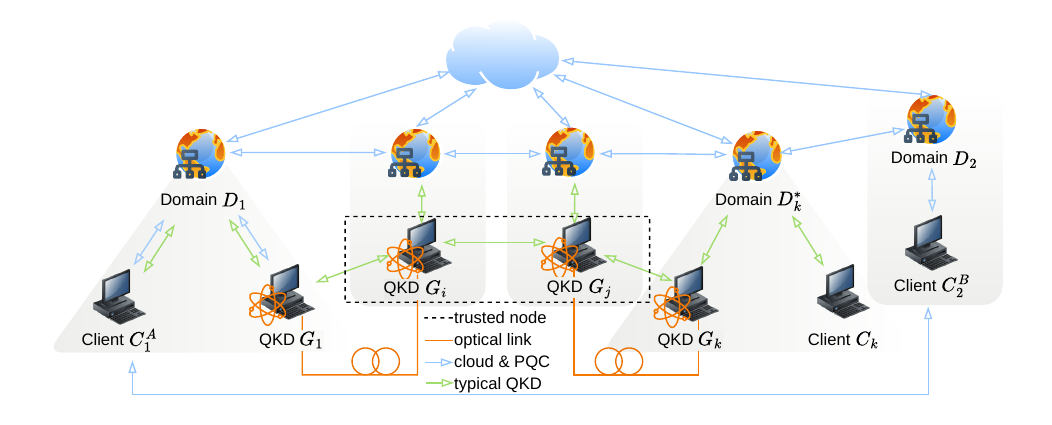}
		\caption{\textbf{The clients are placed in two remote domains, a quantum $D_1$ and a non-quantum $D_2$.} The \textcolor{green!20!gray}{green} arrows represent the conventional QKD possible interactions, which are limited among domains $D_1$, $D_i$, $D_j$, $D^*_k$. The \textcolor{blue!35!white}{blue} arrows represent the extended QKD-derived and quantum-safe key sharing offered by our implementation through forwarding PQC encapsulated messages.}
		\label{fig:interdomain-nonqkd}
	\end{figure}
	
	\subsection{Transfer Phase}
	Assuming that clients $C^A_1$ of domain $D_1$ and $C^B_2$ of domain $D_2$ need to share the private symmetric key $K^{A,B}_{1,2}$ and that key generators $G_i$ of domain $D_i, \ i\in [n]$ and $G_j,\ j\in [n]$ of domain $D_j$ are QKD terminals, we distinguish the following cases depending on their relation: 
	\begin{itemize}	
		\item Key generators $G_1 \equiv G_i$ and $G_2 \equiv G_j$ are a QKD pair, as in Figure~\ref{fig:interdomain}. The DSSP of the source domain $D_1$ acts as a trusted proxy by masking the relay key $K^{A,B}_{1,2}$ with a second QKD-generated key $K'_{1,2}$ using the operation $K^{A,B}_{1,2}\oplus K'_{1,2}$. The result is transmitted to the adjacent domain, which then reconstructs $K^{A,B}_{1,2}$ using the corresponding shared QKD key material.   
		
		\item Key generators $G_1$ and $G_2$ are parts of QKD pairs connected through a TN comprising terminals $G_i$, $G_j$, as in Figure~\ref{fig:interdomain-tn}. The DSSP of the source domain $D_1$ masks the relay key $K^{A,B}_{1,2}$ with a locally shared QKD key $K_1$ using $C_1 = K^{A,B}_{1,2}\oplus K_{1,i}$ and forwards the result through the intermediate domains; the trusted node recovers the original key as $K^{A,B}_{1,2} = C_1 \oplus K_{1,i}$. The trusted node then establishes a new QKD-derived key $K_{j,2}$ with the next domain, computes $C_2 = K^{A,B}_{1,2} \oplus K_{j,2}$, and forwards $C_2$. The destination domain $D_2$ will recover the relayed key with $K^{A,B}_{1,2} =  C_2\oplus K_{j,2}$.  
		
		\item Key generators $G_1 \equiv G_i$ and $G_j$ are connected in one of the ways described in the previous cases and client $C_2$ belongs to a non-quantum domain, as in Figure~\ref{fig:interdomain-nonqkd}. The DSSP in the source domain first computes $C_1 = K^{A,B}_{1,2} \oplus K_{1,i}$ and forwards $C_1$ to the first trusted node, which recovers the original key as $K^{A,B}_{1,2} = C_1 \oplus K_{1,i}$. The trusted node then establishes a new QKD-derived key $K_{j,2}$ with the next domain, computes $C_2 = K^{A,B}_{1,2} \oplus K_{j,2}$, and forwards $C_2$. The DCE recovers $K^{A,B}_{1,2} = C_2 \oplus K_{j,2}$ and, instead of applying another XOR-based masking operation, encapsulates $K^{A,B}_{1,2}$ using a KEM public key associated with the destination domain, producing a ciphertext $C_{\mathrm{KEM}}$ that is transmitted to the destination, where $K^{A,B}_{1,2}$ is recovered through KEM decapsulation.
		
		\item Key generators $G_i$ and $G_j$ are connected in one of the ways described in the previous cases and clients $C_1$, $C_2$ belong to a domains without a QKD terminal. The source domain encapsulates the session key $K^{A,B}_{1,2}$ using the destination domain's KEM public key, producing a ciphertext $C_{\mathrm{KEM}} = \mathrm{Encaps}(pk_{\mathrm{dst}}, K^{A,B}_{1,2})$, which is transmitted across the network. Upon reception, the destination domain recovers the key by performing decapsulation, $K^{A,B}_{1,2} = \mathrm{Decaps}(sk_{\mathrm{dst}}, C_{\mathrm{KEM}})$, enabling secure end-to-end key delivery without relying on QKD-generated key material.
	\end{itemize}
	
	Using the proposed system, the QKD-derived key material is encapsulated with PQC KEMs and exported under cloud supervision through the described transaction-based workflow. While information-theoretic security is inherently confined to the QKD segment, this approach preserves quantum resistance across domain boundaries and enforces controlled access, traceability, and policy compliance. 
	

	

	\section{Architectural Planes and Main Components}\label{sec:components}
\begin{figure}[H]
		\centering
		\includegraphics[width=.6\textwidth]{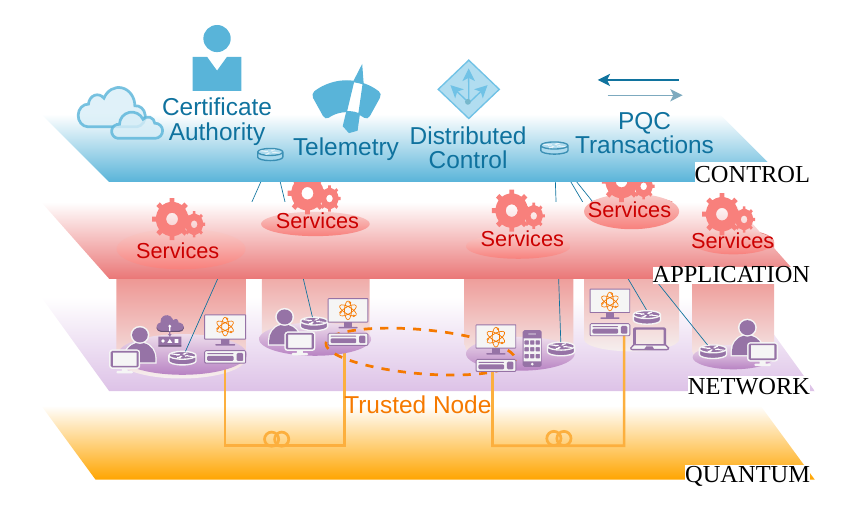}
		\caption{\textbf{Overview of the proposed framework's overall organisation.} The main implementation components are distributed according to the physical layout in the network. We distinguish four (4) planes -- quantum, network, application and control. \textit{Quantum:} defines the quantum channels; \textit{network:} models domains of clients and key generators while specifying TN connections; \textit{application:} outlines service area of action; \textit{control:} distributes crucial management operations among the corresponding cloud services.}
		\label{fig:overview2}
	\end{figure}

	The presented system follows a layered architectural model consisting of an administrative plane, an SDN-orchestrated control plane, and a data plane, spanning the control, application, network, and quantum layers shown in Figure~\ref{fig:overview2}, the latter being treated separately due to its distinct management and operational requirements. The administrative plane constitutes the management of these layers, defining policies, supervising configuration, and controlling security-critical operations. The SDN-orchestrated control plane maintains a global view of the network and enables dynamic configuration and reconfiguration of network paths in response to topology changes or policy decisions. The data plane is responsible for the execution of communication and processing functions, including the transport of user data, the operation of communication channels, and the application of cryptographic mechanisms. Together, these planes provide a clear separation between governance, decision-making, and execution across the system.
	
	\subsection{Administration and Control}
	The administrative plane supervises the configuration and coordination of system components, without directly participating in their operation. It defines and enforces system-wide policies, manages client permissions and possible actions, and regulates the sources of authorization material, including certificates, public key pairs, and tokens, according to user profiles. These services are domain-specific and supervised by the cloud control plane. Cloud services reside alongside the administration control and support system-wide operations, while the administrative plane manages actions with significant security concern, such as client addition and profile creation. The control plane leverages SDN control and automation to dynamically configure network paths, isolate traffic, and react to topology or policy changes.  As QKD networks scale, SDN features that allow dynamic configurations and regulated key retrieval become critical~\cite{mendez_switching_2026}. In our deployment, a visualised SDN provides automated switching for QKD terminals while abstracting the QKD layer. 
	
	\subsection{Certification, Authentication, Identification and Trust}
	Authentication is achieved with a strict scheme that involves certificates, PQC KEM key pairs (asymmetric) and policy alignment (e.g domain specific user access, RBAC, etc.). Client permissions and possible actions are explicitly defined and authorization material (certificates, public key pairs, tokens) sources are restrained according to user profile. The proposed authentication framework is based on a multi-level public key infrastructure (PKI) managed from the cloud, that employs separate roots for QKD nodes of different vendors and another root for the rest of the testbed. In more detail, the overall system employs three (3) root Certificate Authorities (CAs), and four (4) Intermediate Certificate Authorities (ICAs). Separate roots not only ensure compatibility with the QKD nodes, which require distinct chain-of-trust formats, but also preserve integrity among chains and ensure interoperability. Certificates are categorized according to their operational role: 
	\begin{itemize}
		\item infrastructure identity certificates for device-level identification, 
		\item service device-bound identification and authorization  certificates for client-service interactions, and 
		\item key management trust certificates for secure inter-KMS communication and synchronization. 
	\end{itemize}
	Certificate issuance, renewal, revocation, and policy enforcement are managed by dedicated services operating under cloud supervision. Certificate maintenance guarantees dependency dismantling on revocation events; in a partial certificate renewal case wherein a certificate becomes invalidated, the old certificate is returned to re-establish the client or domain, reconstructing the part of the network that could potentially accommodate an adversary; otherwise maintenance triggers system overhauling. 
	
	Trust is established through explicit cross-validation of service certificates rather than reliance on a single centralized issuer. During domain build, gateways are provisioned with certificates used for mutual TLS (mTLS) authentication and clients are defined by the administrator and securely initialised with digital certificates signed with post-quantum algorithms. Any NEs participating in authenticated interactions are also provided a key pair produced with PQC KEMs, bound to their certified identities, and each activated public key $pk^{KEM}$ is stored on the cloud for peer-to-peer identification and trust establishment. Following initial authentication, the system employs proof-of-possession mechanisms to avoid repeated transmission of long-term credentials. Instead, short-lived cryptographic tokens are issued and used as digital tickets for subsequent transactions. Possession of a valid token cryptographically demonstrates authorization, enabling stateless authentication at the service level while ensuring scalability, replay protection, and minimal exposure of sensitive credentials. 
	
	\subsection{Key Management}\label{sec:key-management}
	The key management module orchestrates domain deployment and, as such, determines the behaviour of the final system with respect to trust and the cryptographic material's lifecycle. In more detail, it coordinates the distribution of domains built beforehand and ensures that they are provisioned with certificates and key pairs sufficient to initiate trusted connections. It distributes and maintains certificates for each NE in a domain and connects each NE to a rotating key pair created with the PQC KEM NTRU~\cite{Hoffstein1998NTRU}. Eventually, the KMS provides NEs with an initial transactional KEM pair $(pk^{KEM}, sk^{KEM})$ from which the public key is stored on the cloud for client discovery. Another pair is ephemeral and updated after each transaction with the help of the domain services, while the last is used for identification. Each sender holds a separate public key for each receiver, making sure that the message can be decapsulated and decrypted by the holder of the corresponding secret key, hence the KEM pair is used as an additional identity validation point. 
	With regards to key retrieval, our ZTP device driver architecture ensures automatic provisioning with full agnostic nature. Key generators are provisioned to the DSSP over a specific secondary Network Interface Controller. This physical isolation provides protected access and identification, as each key generator can be registered to the system and identified via different types of identity claims (e.g. PUF, Printed Circuit Board specific features). The DSSP backend, then, uses a key generator-specific driver that handles the key requests using the appropriate security claims. This scheme leads to interoperability and secure assimilation of any type of device for key generation.
	
	\subsection{Vendor-Agnostic QKD Management}
	A significant portion of effort addressed the seamless incorporation of the QKD pairs. QKD management is performed from the cloud and is designed to accommodate the requirements of the devices in our testbed, while remaining readily extensible to support additional hardware. This module is meant to efficiently provide access to cryptographic material while maintaining smooth operation and communication with the domain, as commercial QKD devices frequently depend on rigid configuration requirements, limiting seamless integration in large-scale, multi-vendor environments. In our implementation, we found misalignment on four (4) levels, including administration or management, authentication mechanisms, SDN support and key retrieval interfaces.
	
	
	\subsection{Orchestration} 
	The orchestration layer is responsible for the coordinated management and automation of network, control, and quantum resources across domains. It provides consistent configuration, deployment, and supervision of complex scenarios, but also integrates with the SDN controller to allow dynamic configuration, monitoring, and reconfiguration of network paths in response to topology changes or policy decisions. In the context of this work, orchestration plays a central role in the controlled bridging of high-level components with low-level system elements. A key function of the orchestration layer is the support for automated scenario building, whereby network topologies, domain relationships, and service dependencies can be instantiated programmatically. This includes the definition of participating network elements, inter-domain links, and associated control policies. According to the system operation described in Section~\ref{sec:overview}, three main automated scenarios are prepared:
	\begin{enumerate}
		\item  \textbf{QKD response time:} This test involves all the processes required for the key agreement between a QKD node pair located in separate domains, as well as the retrieval and transmission of its identifier as an encapsulated message across domains. The development of this test allowed to study the rate of key acquisition in a real-world deployment within the context of PQC-protected interdomain connections and revealed the issue of depletion, as some QKD nodes have limited key generation throughput.
		\item \textbf{Intradomain key retrieval:} This test measures the end-to-end latency of service requests, decomposed into the communication and processing stages between user, device, domain, and key generation entities, including the return path of the generated key material. It captures both communication and processing delays involved in request propagation and key delivery.
		\item \textbf{Cloud key retrieval:} This test measures the latency of client-domain-cloud exchanges during service enrollment and token handling. It includes client authentication and the transmission and retrieval of tokens and identifiers via domain services and cloud infrastructure.
	\end{enumerate}
	
	\subsection{Telemetry}
	A major concern since the early stages of the system's development has been the design of an elaborate telemetry component that offers transparency, fault detection and localization, in addition to close performance monitoring. Telemetry has been tuned to be device-oriented and to provide a clear image of the network's status, enabling observability and the rapid identification of problematic segments. To this end, telemetry data are collected and aggregated across domains, allowing both localized and system-wide conditions to be assessed in a consistent manner. The testbed logging has been tuned to avoid obstruction of normal operation, introducing only negligible performance impact while delivering meaningful and actionable updates. Given the high volume of telemetry and event messages generated by the system, device-oriented filtering and aggregation mechanisms have been employed to surface only significant state changes and relevant events, thereby preventing information overload. At the same time, clients are provided with control over telemetry and logging behavior, allowing them to enable, disable, or trigger logging operations manually according to their specific requirements. Logging messages produced cover key acquisition latency, transaction completion, client authentication events, and cryptographic operation timings, such as encryption and decryption. These metrics support both real-time monitoring and post-experiment analysis, contributing to validation, debugging, and performance characterization of the deployed system. 
	
	\section{Experimental results}
	
	We conducted experiments on a testbed of ten (10) domains containing QKD nodes from three vendors --Toshiba, ID Quantique, ThinkQuantum-- alongside two (2) domains without QKD nodes (Figures~\ref{fig:topology}). Each domain includes a gateway router, a low-power DSSP device, and client nodes (Table~\ref{tab:devices}), while an aggregation router enables inter-domain and cloud connectivity via an SDN-enabled data plane. QKD communication relies on ETSI 014~\cite{ETSI014}.
	
	Connection latency is between 10 and 25 $\mathrm{ms}$ per link, with the multi-authentication combination mTLS and PQC KEM included in the measurement. SDN functionality  increases latency by a $\sim10\ \mathrm{ms}$ SDN overhead. Intradomain transactions complete in $t \in [10 + x, 25 + x]\ \mathrm{ms}$, where $x \in \{20, 250\}\ \mathrm{ms}$ depending on the QKD vendor, while cloud access introduces an overhead of $\sim13\text{ to }52\ \mathrm{ms}$ per request. The dominant latency factor is QKD key retrieval, which is approximately one order of magnitude slower than classical cryptographic and networking operations, whereas network jitter remains negligible. Table~\ref{tab:enc-dec} reports the processing overhead introduced by encapsulation and decapsulation operations across heterogeneous DSSP platforms.
	
	For PQC, NTRU761 KEM pair generation -- measured using a fully portable version of the algorithm without processing optimisation instructions -- requires $37.6\ \mathrm{ms}$ per device on a laptop and $71\ \mathrm{ms}$ in cloud containers, with per-domain generation (four pairs) reaching $125\ \mathrm{ms}$ and $280\ \mathrm{ms}$ respectively. Increasing CPU cores (2 vs. 10) does not improve latency, indicating dependence on single-core performance. PQC operations remain stable across heterogeneous environments.
	
	\begin{table}[h]
		\centering
		\begin{tabular}{lc@{}c}
			\toprule
			\bfseries Platform &\bfseries Role &\bfseries Specifications \\\midrule
			Poweredge R730 & cloud servers & 32 cores 32GB Ubuntu server 24.04 PCI exp 3.0 \\
			Laptop & client & I713700H 20 Cores 16GB Ubuntu 25.04 PCIE 4.0 \\
			Raspberry Pi 5 & DSSP  & 2GB 4 core Aarch64 kernel SMP  \\
			Raspberry pi 4 & DSSP  & Raspbian lite 32GB heatsink no fan 6.12 SMP PREEMPT \\
			RISCV64		   & DSSP  & Sifive P550 premier Arm RISCV64 32GB SMP PREEMPT\_DYN \\\bottomrule
		\end{tabular}
		\caption{Devices used in our testbed with their specifications and role.}
		\label{tab:devices}
	\end{table}
	
	\begin{table}[h]
		\centering
		\begin{tabular}{l@{}c@{}c@{}}
			\toprule
			\bfseries Platform &\bfseries Encryption &\bfseries Decryption \\\midrule
			RISCV64 (full desktop)      & 2.5 ms     & 24 ms      \\
			Raspberry Pi 4 (full desktop)    & 2.5 - 4 ms & 6.5 ms     \\ 
			Raspberry Pi 5 (no desktop) & 2.5 ms     & 6.5 ms     \\
			\bottomrule
		\end{tabular}
		\caption{Times for encryption and decryption of the encapsulated packet travelling between domains.}
		\label{tab:enc-dec}
	\end{table}
	
	\begin{table}
		\centering
		\begin{tabular}{l@{}cccc}
			\toprule
			\bfseries Platform & \bfseries 1 KEM pair & \bfseries 4 KEM pairs & \bfseries CA - ICA (18 crts) & \bfseries QKD (16 crts) \\\midrule
			Laptop & 37.58 & 125.26 & 411.82 & 1227.35 \\
			Cloud server (2 cores) & 71 & 280.31 & 4377.59 & 16888.82 \\
			Cloud server (10 cores) & 71 & 280.31 & 4377.59 & 13443.88 \\\bottomrule
		\end{tabular}
		\caption{Measurements regarding the creation of key pairs and certificates on each device of Table~\ref{tab:devices}. The KEM used is NTRU761, while a part of the listed certificates (crts) were produced with  ECDSA P-384 and the rest with RSA 4096. the results are reported in milliseconds.}
		\label{tab:results}
	\end{table}
	
	Certificate generation is dominated by RSA-4096 costs. On the laptop, CA-ICA tree construction (18 certificates) requires $412\ \mathrm{ms}$, while QKD infrastructure initialization takes $\sim1227\ \mathrm{ms}$, with only $\sim27\ \mathrm{ms}$ attributed to ECDSA P-384 and $\sim1200\ \mathrm{ms}$ to RSA. In cloud environments, initialization increases to $\sim13.4\text{--}16.9\ \mathrm{s}$, again largely due to RSA generation, with minimal gains from additional CPU cores. Table~\ref{tab:results} summarizes the measured certificate and KEM generation overhead across platforms.
	
	Overall, system overhead from SDN control and service orchestration remains low and manageable even on constrained devices. In contrast, QKD key retrieval and infrastructure initialization constitute the primary bottlenecks, with vendor-specific implementations introducing higher latency in exchange for robustness and advanced authentication support. Classical certificate generation is resource-sensitive, whereas PQC operations exhibit consistent performance across platforms.
	
	\section{Conclusions}
	
	The presented work demonstrates a flexible multi-domain quantum-secure network architecture that combines QKD, PQC, SDN orchestration, and cloud-managed trust services into a unified framework. The proposed approach accommodates heterogeneous deployments, including fully PQC-based domains, hybrid quantum-classical infrastructures, and multiple types of key generators through an agnostic ZTP-driven integration model. This allows secure key transport and authenticated communication across domains without requiring homogeneous QKD deployments or tightly coupled vendor-specific infrastructures. The implementation further highlights the practical challenges associated with the integration of multi-vendor QKD systems, particularly with respect to incompatible management interfaces, authentication schemes, orchestration mechanisms, and key retrieval procedures. At the operational level, the limited key streaming and generation rates of current QKD nodes remain a major bottleneck compared to the relatively stable overhead introduced by PQC operations and SDN functionality. Nevertheless, the measurements indicate that the additional orchestration, authentication, and encapsulation layers can be integrated with limited performance impact even on constrained devices, supporting the practicality of the proposed architecture in realistic deployments.	Another important aspect of the proposed framework is its adaptability. The architecture can accommodate a wide range of deployment scenarios, including domains with native QKD support, partially connected trusted-node chains, and purely classical domains relying exclusively on PQC mechanisms for secure key transport. At the same time, the DSSP abstraction and cloud-supervised management model provide a consistent operational interface independently of the underlying cryptographic or networking technologies, simplifying interoperability and service orchestration across heterogeneous environments. Future work will focus on extending the infrastructure towards geographically remote domains and larger-scale deployments involving significantly higher numbers of clients and services. Such scaling introduces additional challenges related to orchestration complexity, latency accumulation, inter-domain trust propagation, certificate lifecycle management, QKD key depletion, telemetry aggregation, and the scalability of centralized cloud and SDN control services. Nevertheless, the modular and vendor-agnostic design of the proposed architecture provides the flexibility required to progressively address these challenges while enabling the evolution towards practical large-scale quantum-secure networks.
	
	\bibliography{references}
	
	\newpage
	\begin{appendices}

		\section{Hardware and Devices}
		
		This section contains photographs of key devices used for the deployment of the enterprise-grade network.
		
		\subsection{Cloud Servers}
		
		\begin{figure}[H]
			\centering
			\includegraphics[height=.47\linewidth, angle=270]{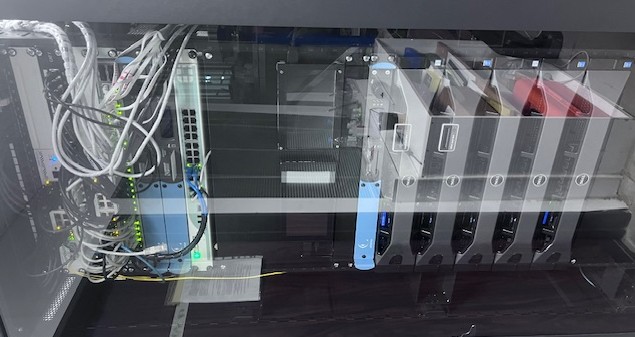}
			\caption{The cloud servers}
		\end{figure}
		
		\subsection{QKD Nodes}
		
		\begin{figure}[H]
			\centering
			\includegraphics[width=.7\linewidth, angle=0]{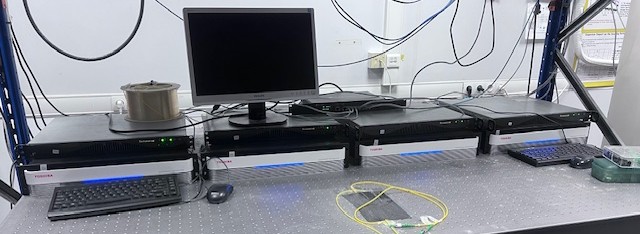}
			\caption{Toshiba MU nodes}
		\end{figure}
		
		\begin{figure}[H]
			\centering
			\includegraphics[height=.7\linewidth, angle=90]{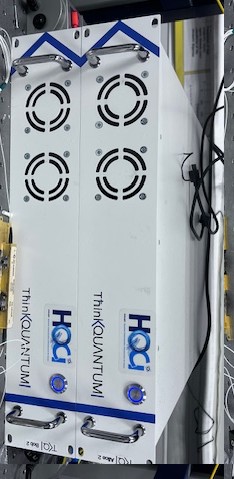}
			\caption{ThinkQuantum QUKY}
		\end{figure}
		
		\begin{figure}[H]
			\centering
			\includegraphics[height=.7\linewidth, angle=270]{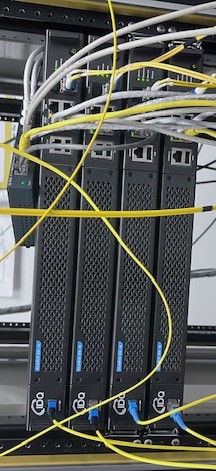}
			\caption{ID Quantique Clavis XG}
		\end{figure}
		
		\section{Additional Figures}
		
		\subsection{Trusted Node Emulation Through MACsec}
		
		\begin{figure}[H]
			\centering
			\includegraphics[width=\textwidth]{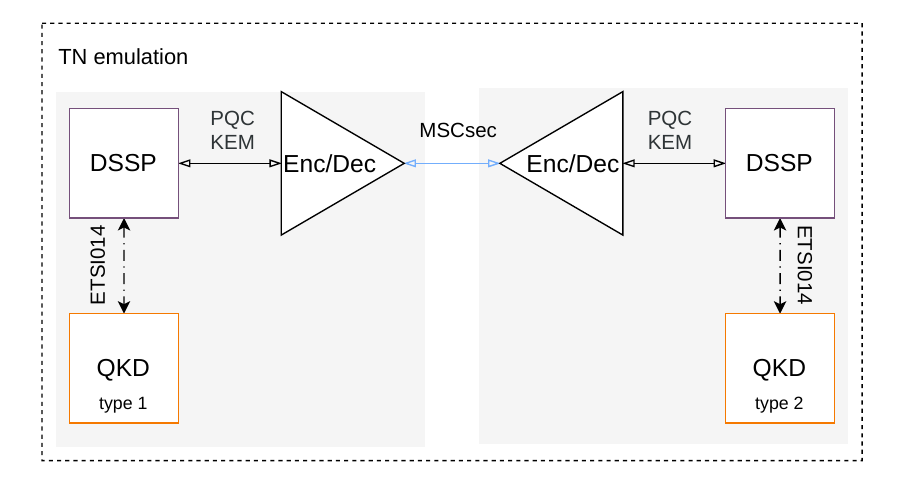}
			\caption{
				Emulated trusted-node operation through MACsec. The DSSPs establish PQC-protected tunnels and relay QKD-derived key material across domains.
			}
			\label{fig:dssp_tn}
		\end{figure}
		
		\subsection{QKD Chain with DSSP Relay}
		
		\begin{figure}[H]
			\centering
			\includegraphics[width=\textwidth]{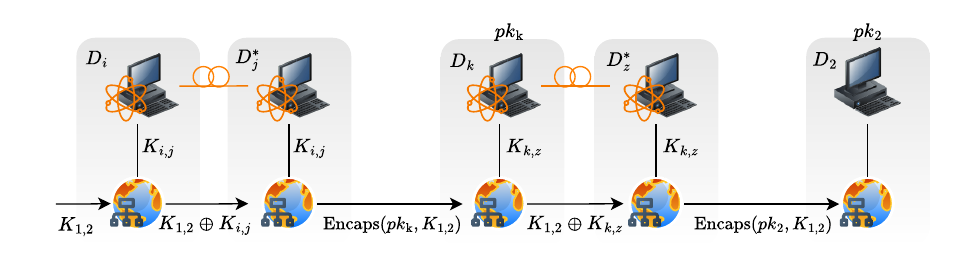}
			\caption{
				Example of a chained QKD relay across multiple domains. Intermediate DSSPs
				recover and re-protect the relayed key material before forwarding it to the
				next domain.
			}
			\label{fig:qkd_chain}
		\end{figure}
		
		\newpage
		\section{Comparison with Similar Works}
		
		\begin{table*}[h]
			\centering
			\scriptsize
			\setlength{\tabcolsep}{3pt}
			\renewcommand{\arraystretch}{1.15}
			\caption{Comparison of the ETSI/KMS federation architecture~\cite{barral_interconnecting_2026}, the MadQCI SDN-QKD deployment~\cite{martin_madqci_2024, mendez_sdn-based_2024}, and the proposed DSSP-based multi-domain framework.}
			\label{tab:comparison_extended}
			\begin{tabularx}{\textwidth}{|p{2.4cm}|X|X|X|X|}
				\hline
				\textbf{Aspect} &
				\textbf{ETSI/KMS Federation} &
				\textbf{MadQCI SDN-QKD} &
				\textbf{MadQCI Hybrid Intercommunication} &
				\textbf{Proposed DSSP Framework} \\
				\hline
				
				Primary objective &
				Federation of heterogeneous regional QKD networks through ETSI-compliant KMS overlays. &
				Deployment of scalable SDN-QKD infrastructure integrated into production telecom networks. &
				Hybrid quantum-safe inter-domain communication using SDN, QKD, and PQC across administrative domains. &
				Enterprise-grade quantum-secure orchestration integrating QKD, PQC, SDN, and cloud trust services. \\
				\hline
				
				Architectural paradigm &
				Distributed KMSTN federation overlay. &
				Software-defined QKD networking (SDN-QKD). &
				SDN-based hybrid QKD/PQC domain intercommunication. &
				Domain-oriented DSSP orchestration framework. \\
				\hline
				
				Core orchestration entity &
				KMS / KMSTN federation nodes. &
				SDN controller with LKMS and forwarding modules. &
				SDN controllers coordinating LKMSes and border nodes across domains. &
				Distributed DSSPs supervised by cloud orchestration. \\
				\hline
				
				Relay abstraction &
				Trusted KMS relay proxies. &
				Trusted relay nodes with forwarding modules separated from LKMS. &
				Border-node relay architecture with SDN-assisted key forwarding. &
				DSSPs acting as trusted inter-domain relay proxies. \\
				\hline
				
				Deployment model &
				Interconnection of regional QKD islands. &
				Production deployment across operational telecom facilities and multiple operators. &
				Interconnection of heterogeneous administrative and technological domains within MadQCI. &
				Enterprise/cloud multi-domain deployment with SDN-managed orchestration. \\
				\hline
				
				Network domains &
				Federated regional domains. &
				Telefónica and RedIMadrid domains interconnected through border nodes. &
				Cross-domain QKD forwarding between Telefónica, RedIMadrid, and Munich domains. &
				Multiple enterprise or carrier domains interconnected through DSSP federation. \\
				\hline
				
				QKD technologies &
				Hybrid QKD/PQC federation. &
				Simultaneous DV-QKD and CV-QKD deployment. &
				Hybrid QKD forwarding with PQC-assisted long-haul emulation. &
				QKD-agnostic integration with optional PQC-only segments. \\
				\hline
				
				Vendor interoperability &
				ETSI-compliant interoperability among QKD regions. &
				Integration of Huawei, Toshiba, ID Quantique, AIT, ADVA, R\&S, and QuSide systems. &
				Vendor-independent inter-domain hybridization using ID Quantique and Huawei systems. &
				Vendor-agnostic orchestration integrating Toshiba, IDQ, ThinkQuantum, and future vendors. \\
				\hline
				
				Control plane &
				Secure KMS coordination and routing. &
				SDN controller implementing routing, QoS, switching, and orchestration. &
				Independent SDN controllers coordinating secure inter-domain key forwarding. &
				Cloud-supervised SDN orchestration integrated with DSSP policy management. \\
				\hline
				
				Switching and path management &
				Logical key-routing federation. &
				Dynamic optical switching enabling 45 direct quantum paths. &
				Dynamic SDN-assisted key forwarding and border-node path orchestration. &
				Dynamic SDN-assisted inter-domain relay selection and routing. \\
				\hline
				
				Quantum/classical coexistence &
				Focused on federation and secure transport. &
				Quantum and commercial telecom traffic coexist on shared fibers. &
				QKD domains interconnected through PQC-emulated long-haul links over classical infrastructure. &
				Supports hybrid quantum/classical infrastructure and DCE extension. \\
				\hline
				
				PQC integration &
				Kyber-secured transport and hybrid relay. &
				Hybrid operation with QKD and conventional cryptographic fallback. &
				PQC KEMs (Kyber/NTRU) used for long-distance QKD emulation and hybrid XOR-based protection. &
				PQC integrated into identity, authentication, orchestration, and secure tunnels. \\
				\hline
				
				Hybridization method &
				Hybrid relay using QKD/PQC-derived secrets. &
				Parallel coexistence of QKD and classical cryptography. &
				XOR hybridization of multiple QKD links and PQC-derived keys at border nodes. &
				QKD-assisted relay with PQC-secured DCE bridging and orchestration. \\
				\hline
				
				Standards alignment &
				ETSI GS QKD 014 and 020 oriented. &
				ETSI GS QKD 004, 014, and 015 compliant. &
				Strong ETSI GS QKD 004/014/015 alignment with SDN integration. &
				ETSI GS QKD 014/015 aligned with extensible DSSP APIs. \\
				\hline
				
				Key management &
				Distributed inter-domain KMS federation. &
				LKMS coordinated by SDN controller. &
				Distributed LKMSes with SDN-assisted inter-domain key forwarding. &
				DSSP-centric orchestration with cloud-assisted trust services. \\
				\hline
				
				Long-distance support &
				PQC-secured inter-domain federation. &
				Metropolitan-scale SDN-QKD deployment. &
				PQC-emulated long-haul QKD links between Madrid and Munich. &
				Supports geographically distributed DCE-connected domains. \\
				\hline
				
				Operational philosophy &
				Carrier-grade federated KMS architecture. &
				Production-grade telecom-integrated QKD networking blueprint. &
				SDN-enabled hybrid quantum-safe inter-domain communication framework. &
				Software-defined enterprise quantum-secure networking framework. \\
				\hline
				
				Main contribution &
				Hybrid PQC/QKD inter-domain federation using ETSI-compliant KMS overlays. &
				Large-scale heterogeneous SDN-QKD deployment in real telecom infrastructure. &
				Practical SDN-based inter-domain QKD/PQC forwarding and border-node hybridization. &
				Vendor-agnostic DSSP framework for practical multi-domain quantum-secure orchestration. \\
				\hline
				
			\end{tabularx}
		\end{table*}
	\end{appendices}
	
	\end{document}